\documentclass{article}      
\usepackage{graphicx}
\usepackage{epstopdf}
\title{Quark Model Calculations of the N to Delta Reaction}

\author{M.M. Giannini\\
Dipartimento di Fisica dell'Universit\`a di Genova\\
and \\
Istituto Nazionale di Fisica Nucleare, Sezione di Genova, Italy \\
}

\date{}
\begin{document}

\maketitle  

\begin{abstract}
The electromagnetic excitation of the nucleon resonances is studied in the
framework of Constituent Quark Models. Particular attention is devoted to
the transition to the $\Delta$ resonance and to the issue of a possible 
deformation of hadronic systems, mainly in connection with the problem of
the quadrupole excitation. The analysis of the emerging discrepancies
between data and theoretical predictions is discussed and shown to lead to
important conclusions concerning the internal dynamical structure of
hadrons.

\end{abstract}

\section{Introduction}

The study of the electromagnetic excitation of the nucleon resonances is
expected to provide a good test for our knowledge concerning the internal
structure of baryons. The $N-\Delta$ transition is one of
the most important ones, since both in the pion-nucleon and the
electroproduction channels the $\Delta$ peak is the most evident and
lowest one in energy. Moreover, according to the quark model, the
$\Delta$ state is the $SU(6)$ partner of the nucleon and therefore, apart
from a spin-isospin flip, it shares the same internal structure with the
nucleon. From a fundamental point of view, the description of the
nucleon resonances and their excitation should be performed within a QCD
approach, at least in its Lattice formulation. There is considerable 
progress in this area (see e.g. \cite{ND_lat}), however a complete and
consistent account of the whole excitation spectrum is not yet available. 

In this respect, Constituent Quark Models (CMQs) are
particularly useful, since they allow to take into account
fundamental aspects of the quark dynamics within a simple scheme, which
can on the other hand be used for a systematic and consistent study of
various baryon properties.  

In the following, some of the more popular CQMs will be briefly reviewed
and compared, showing to which extent they have been applied to the study
of a
large variety of hadron properties. The electromagnetic (e.m.) excitation
of the nucleon resonances will be discussed, with particular
attention to the $N-\Delta$ problem. The CQMs permit a systematic
comparison between the theoretical predictions and the observed
experimental behaviour and are able to describe in a
quantitative way many important data. The analysis of the emerging
discrepancies allows then to draw reasonable conclusions
concerning some fundamental 
aspects of the quark dynamics, such as meson and/or sea-quark effects,  
which are presently missing, but expected to play a relevant role in the
future of hadron physics.

\section{Constituent Quark Models}

Various Constituent Quark Models (CQM) have been proposed in the past
decades after the pioneering work of Isgur and Karl (IK) \cite{ik}. Among
them
let us quote the relativized Capstick-Isgur model (CI)
\cite{ci}, the algebraic approach (BIL) \cite{bil}, the hypercentral CQM
(hCQM) \cite{pl}, the chiral Goldstone Boson Exchange model ($\chi$CQM)
\cite{olof} and the Bonn instanton model (BN) \cite{bn}. They are all able
to reproduce the baryon spectrum, which is the first test to be performed
before applying any model to the description of other physical quantities.
The ingredients of the models are quite different, but they have a simple
general structure, since the interaction $V_{3q}$ they use can be split
into a
spin-flavour independent
part $V_{inv}$, which is $SU(6)$-invariant and contains the confinement
interaction, and a $SU(6)$-dependent part $V_{sf}$, which contains spin
and eventually flavour dependent interactions

\begin{equation} \label{v3q}
V_{3q}�=�V_{inv}�+�V_{sf}
\end{equation}

This structure should be compared with the
prescription provided by the early Lattice QCD calculations \cite{deru},
that is an interaction containing a spin-indepent long range part and a
spin-dependent short range term. The pattern of Eq. (\ref{v3q}) is
actually in agreement with an important
feature of the baryon spectrum. In fact, the various resonances can be
grouped into
$SU(6)$-multiplets, the energy differences within each multiplet being at
most of the order of $15\%$ (as in the case of $N-\Delta$ mass difference 
and of the splittings within the $SU(3)$ multiplets). To illustrate this
point, 
we list below the lower
$SU(6)$-multiplets accompanied by the non strange baryons they contain:

$
\begin{array}{llllcllll}
\\
\\
& & & & (56, 0^+): & P_{11}(938) & P_{33}(1232) & & \\
\\
& & & & (56*,0^+): & P_{11}(1440) &  P_{33}(1600) & & \\
\\
& & & & (70, 1^-): & D_{13}(1520) & S_{11}(1535) & S_{31}(1620) &
S_{11}(1650)
\\
\\
& & & &           & D_{15}(1675) & D_{33}(1700) & D_{13}(1700) & \\
\\
& & & & (56, 2^+): & F_{15}(1680) & P_{13}(1720) & F_{35}(1910) &
P_{33}(1920) \\
\\
& & & &           & F_{37}(1950) &              &              & \\
\\
& & & & (70, 0^+): & P_{11}(1710) & P_{31}(1910) &              & \\
\\
\\
\end{array}
$

The notation for the $SU(6)$-multiplets is $(N, L^P)$, where $N $ is the
dimension of the $SU(6)$-representation, $ L$ is the total orbital
angular momentum of the three-quark state describing the baryon and $P$ 
the corresponding parity. The star in the second line reminds that the
states have the same spin-isospin structure as those in the first line
but are radially excited. It should be reminded that with three quarks one
can obtain
the $SU(6)$-representations with dimensions $N=20, 56, 70$. The spin and
flavour content of each $SU(6)$-representation is well defined, since
the three $SU(6)$ representations can be decomposed according to the
following scheme

\begin{equation}
20�=�^4 \underline{1}�+�^2 \underline{8} 
\end{equation}
\begin{equation}
56�=�^2 \underline{8}�+ �^4 \underline{10}  
\end{equation}
\begin{equation}
70�=�^2 \underline{1}�+�^2 \underline{8}�+�^4 \underline{8}�+�^2
\underline{10} 
\\
\end{equation}

The suffixes in the r.h.s. denote the multiplicity $2S+1$ of the $3q$ spin
states
and the underlined numbers are the dimensions of the $SU(3)$
representations. This means for instance that the
$56$ representation contains a spin-$1/2$ $SU(3)$ octect and a spin-$3/2$
$SU(3)$ decuplet (more details can be found in \cite{mg}). 

An important observation concerns the level ordering displayed by the
experimental baryon masses: the $1^-$ states are in average almost
degenerate with the first $0^+$ excitation, while for any two body
potential the ordering is $0^+, 1^-, 0^+$. Moreover, in the case of the
h.o. potential the spacing between two shells is the same over the whole
spectrum. 

In order to reproduce the spectrum, any CQM should lead to reasonable
average energy levels by means of the $SU(6)$-invariant part of the
potential $V_{inv}$ and describe the splittings within each multiplet
through the
spin-flavour dependent interaction $V_{sf}$. The latter is relevant for
the
topic
of this workshop, in fact the (spin-dependent) tensor forces,
which are
present in some models, are able to generate a deformation of the three
quark states. It is therefore useful to analyze the main features of the
various models.

CI \cite{ci}. The confinement is provided by a three-body term
corresponding to a $Y-$shaped configuration. The multiplet splittings are
mainly given by an interaction, which is inspired by the
One-Gluon-Exchange
mechanism and as such it contains a spin-spin term and a tensor force. The
three-body equation, with relativistic kinetic energy, is solved by means
of a variational approach in a large h.o. basis.

BIL \cite{bil}. The $SU(6)$-invariant part of the levels is obtained
starting from a  $U(7)$-symmetry of the three-quark states and considering
a string-like collective model for the mass operator of the baryon states,
taking into account rotations and vibrations of a $Y-$shaped
configuration. The energy splittings are produced by a G\"{u}rsey-Radicati
mass formula \cite{gura}, containing constant spin, isospin and flavour
dependent
terms, which are proportional to the Casimir operators of the $SU(6)$,
$SU(2)$ and $SU(3)$ groups describing the relevant intrinsic quark degrees
of freedom.

hCQM \cite{pl}. The quark potential is assumed to be hypercentral, that is
to depend only on the hyperradius $x$, defined as $x�=�\sqrt{\vec{\rho}^2
+\vec{\lambda}^2}$, where $\vec{\rho}$ and $\vec{\lambda}$ are the Jacobi
coordinates describing the quark internal motion. The hyperradius $x$
assumes the meaning of a collective variable, describing the size of the
baryon state. The explicit form of the potential is given by
\begin{equation}
V_{hCQM} = - \frac{\tau}{x} + \alpha x
\end{equation}
A potential containing a coulomb-like and a linear confinement term has
been used since long time in the description of the meson sector (Cornell
potential). Such structure has been recently supported by Lattice QCD
calculations for static quarks \cite{bali,alex,sug}. In this respect the
hCQM
potential can be considered as the hypercentral approximation of a
quark-quark interaction of the Cornell type. The hypercentral
approximation has been used both in the nuclear \cite{ballot,hca} and
baryon
\cite{has} cases, with good results specially for the lower part of the
spectra. Thanks to the $x$-dependence, the hCQM interaction may also
include
many body contributions, corresponding for instance to the already
mentioned $Y$-shaped string configuration. The presence of the
coulomb-like term is important for various reasons. Here it is sufficient
to remind that the potential $1/x$ leads to an analytical solution,
thereby providing an alternative basis to the h.o. one, and moreover the
$1^-$ states are perfectly degenerate with the excited $0^+$ states.
The linear confinement term slightly modifies this ordering
\cite{pl}, and, in order to obtain the correct position of the resonances,
in particular of the Roper $P_{11}(1440)$, it is necessary to add isospin
dependent terms to the potential \cite{iso}. The multiplet splittings are
provided by a hyperfine interaction of the standard form \cite{ik}. The
description of the spectrum has been extended to the strange resonances by
means of a G\"{u}rsey-Radicati $V_{sf}$ term \cite{gr}. The fit of the
spectrum leads to the values $\tau�=�4.59, \alpha�=�1.61�(fm)^{-2}$
\cite{pl}, which are kept fixed in the subsequent applications of the
model to various quantities of interest.

GBE \cite{olof}. The confinement interaction, in the more recent version
of the model, is given by a linear two-body term. Consistently with the
idea that at low energies pseudoscalar mesons are relevant degrees of
freedom as Goldstone Bosons, an explicit quark-quark
potential due to
meson exchange is introduced. The splittings within multiplets are then
provided by the spin and isospin dependence of the pseudoscalar meson
exchange. The model has been extended to include also scalar and vector
meson exchange \cite{ple}. Of course, both $\pi$ and $\rho$ exchange lead
to tensor forces and therefore to a possible deformation of the nucleon
and of the $\Delta$ resonance.

BN \cite{bn}. The model is fully relativistic in the sense that it is
based on a Bethe-Salpeter approach for the description of the three-body
system. The confinement is produced by a three-body term depending
linearly on a collective variable corresponding to a $\Delta-$shaped
three-quark configuration. The result for the spectrum are only
slightly changed if the $\Delta-$shaped
three-quark configuration is substituted with the $Y$ or the hypercentral
one. The $V_{sf}$ part is provided by a
two-body 't Hooft's residual
interaction, based on QCD-instanton effects. It is important to
mention that such interaction acts only on antisymmetric two quark spin
states and therefore it does not affect the $\Delta$ resonance.

As mentioned above, all models provide a more or less reasonable
description of the baryon spectrum; in particular the $N-\Delta$ mass
difference is correctly fitted, although this splitting has various
origins in the different models: hyperfine interaction (GI, hCQM), pion
exchange (GBE), instanton effects (BN), spin-isospin dependent Casimir
operators (BIL). 

The CQMs have also been applied to the calculation of other physical
quantities and it is interesting to see to which extent and how
systematically the various CQM have been used; one should however not
forget that in many cases
the calculations refered to a CQM calculations are actually performed
using a simple h.o. wave function for the internal quark motion either in
the non relativistic (HO) or relativistic (relHO) framework. In
the following a (non exhaustive) list of applications of the
various CQM models (HO, relHO, IK, CI, BIL, hCQM, GBE, BN) is reported:

- photocouplings: HO \cite{cko}, IK \cite{ki}, CI \cite{cap}, BIL
\cite{bil}, hCQM
\cite{aie} (for a comparison among these and other approaches see e.g.
\cite{aie,cr2}); 

- the transition form factors for the excitation of the nucleon resonances
(helicity amplitudes): HO \cite{cko}, KI \cite{ki}, relHO \cite{ck}, CI
\cite{cap,card_ND,card_Roper}, hCQM \cite{aie2,sig1,mds2}, BN \cite{mert},
in the latter case with particular attention to the strange baryons
\cite{cau};

- the elastic nucleon form factors: BIL \cite{bil,bilff}, CI
\cite{card_ff}, hCQM \cite{mds,sig1,rap,ff_05}, GBE \cite{wagen,boffi}, BN
\cite{mert}, again with emphasis on the strange baryons \cite{cau2};

- the axial nucleon form factors GBE \cite{boffi,gloz} BN \cite{mert};

- the strong decay of baryon resonances IK \cite{ki}, relHO \cite{cr3}, CI
\cite{cr}, BIL \cite{bil3}, GBE \cite{melde}, hCQM \cite{bad}.

There also calculations of the nucleon structure functions
\cite{facc} and of the Generalized Parton
Distributions \cite{BPT,sco}, performed using simple CQMs, eventually in a
relativized framework.

Here however
the attention will be devoted to the
$N-\Delta$ transition both in the photon limit and in its full $Q^2$
dependence.

\section{The electromagnetic transition amplitudes}

The attempt of describing the transition amplitudes for the
electromagnetic excitation of the baryon resonances implies a more
stringent test of the dynamics involved in the various CQMs, since this
kind of process is more sensitive to the internal structure of the three
quark states.

From the experimental point of view, the transition amplitudes are
extracted from measurements of the photo- or electro-production of pions.
This leads to a problem concerning the sign of the e.m. amplitude. In
fact, one can determine the overall phase of the pion production
amplitude,
however the sign of the e.m. excitation vertex is strictly correlated to
the one of the strong decay. Presently, CQMs are not able to calculate in
a consistent way the pion production process and it is therefore important
to extract in any case the e.m. amplitude, even if it includes the unknown
phase of the strong decay vertex (for a discussion on this point
see e.g. \cite{cr2}).

In order to calculate the e.m. transition amplitudes within a CQM, one
considers a direct coupling between the quark current and the e.m. field.
The quark current is chosen in the majority of cases as a one-body current
in impulse appproximation. In this approximation, the transverse
photon-quark interaction can be written

\begin{equation} \label{eq:htm}
H^t_{em}~=~-~\sum_{i=1}^3~\left[\frac{e_j}{2m_j}~(\vec{p_j} \cdot
\vec{A_j}~+
~\vec{A_j} \cdot \vec{p_j})~+~2 \mu_j~\vec{s_j} \cdot (\vec{\nabla}
\times \vec{A_j})\right]~~,
\end{equation}
in Eq. (\ref{eq:htm}) $~m_j$, $e_j$, $\vec{s_j}$ , $\vec{p_j}$ and
$\mu_j~=~\frac{e_j}{2m_j}$
denote the mass, the electric charge, the spin, the momentum and the
magnetic moment of the j-th quark, respectively, and
$\vec{A_j}~=~\vec{A_j}(\vec{r_j})$ is the photon field.

The non relativistic formulation of Eq. (\ref{eq:htm}) can be relativized 
introducing various
higher order spin dependent terms (see e.g. \cite{cr2} and references
quoted therein). There are also formulations which make use of a
covariant quark current in the framework of a relativistic dynamics
approach \cite{card_ND}.

The quark-photon interaction of Eq. (\ref{eq:htm}) (or its relativistic
version) is used to calculate the helicity amplitudes for the excitation
of the baryon non strange resonances

\begin{equation}
\begin{array}{rcl}
A_{1/2}(Q^2)&=& \langle B, J', J'_{z}=\frac{1}{2}\ | H^t_{em}| N, J~=~
\frac{1}{2}, J_{z}= -\frac{1}{2}\
\rangle\\
& & \\
A_{3/2}(Q^2)&=& \langle B, J', J'_{z}=\frac{3}{2}\ | H^t_{em}| N, J~=~
\frac{1}{2}, J_{z}= \frac{1}{2}\
\rangle\\
\end{array}
\end{equation}

There are two important aspects in connection with the phenomenological
e.m. helicity amplitudes: the strength at $Q^2=0$ (photocouplings) and the
$Q^2$ behaviour (transition form factors).

As for the photocouplings, the calculations in general describe the
overall
trend, in the sense that they reproduce \cite{cko,ki,bil,aie} the
oscillatory behaviour
displayed by data if increasing masses of 
the $N$ or $\Delta$ resonance states are considered. This means in
particular that the locations of zero or very small strength are accounted
for. However, there is also an equally general underestimate of the strength, 
quite
independently of the model which is used \cite{cko,ki,bil,aie}; the
inclusion of relativistic corrections with spin dependent terms does not
improve the fit (for a discussion see e.g. \cite{aie,cr2}). The
fundamental reason of this common failure is probably due to the fact that
all CQMs have the same spin-isospin structure and, as it will be discussed
later, also because meson or quark pair effects are lacking.
 
The behaviour of the theoretical transition form factors at low $Q^2$ is
affected by the lack of strength mentioned above. For higher $Q^2$, the
results are strictly dependent on the quality of the three-quark wave
function, that is on the dynamics which is chosen. For instance, for wave
functions dominated by pure h.o., the resulting transition form factors
are too strongly damped for increasing $Q^2$ and therefore are not able to
reproduce the phenomenological trend. In the case of the negative parity
non strange resonances, the experimental behaviour is fairly well
reproduced by the parameter-free calculations made with the hCQM
\cite{aie2}, specially for the $S_{11}$, whose trend has been predicted
before the recent Jlab data \cite{burk02}. The softer $Q^2$ dependence
displayed by the theoretical form factors is due, in the hCQM calculation,
to the presence of the Coulomb-like term in the interaction. In fact,
very similar results are obtained in the analytical version of the model
\cite{sig1}, in which the linear confinement is treated as a perturbation
and the (analytical) $Q^2$ dependence is completely determined by the
$1/x$ potential \cite{sig1}.

Coming to the $N-\Delta$ excitation, the main $M1$ transition can be
evaluated in all models, but the theoretical amplitudes underestimate the
phenomenological one by a factor of the order of $30 \%$; this feature was
present also in the early quark model calculations, which gave the result
\cite{blp,bm} 
\begin{equation}  \label{eq:M1}
G_{M1}�=�\frac{2�M_{\Delta}}{M�+�
M_{\Delta}}�\frac{2}{\sqrt{3}}�\mu_P�=�3.7 
\end{equation} 
where  $M$ and $M_{\Delta}$ are the nucleon and $\Delta$ mass,
respectively and
$\mu_P$ is
the proton magnetic moment; the value in Eq. (\ref{eq:M1}) should be
compared with the measured one, which is of the order of $5$. The
situation has not
been modified by the more recent and refined models.

The $E2$ transition is important because it is connected with the issue
of a possible deformation of hadrons. If the quarks in the Nucleon and the
$\Delta$ are in a pure $S-$wave state there is obviously no $E2$
excitation \cite{bm}. Therefore a deformation can be produced only if the
interaction contains a tensor force: this happens in models with a
hyperfine interaction inspired by QCD \cite{ikk,ci,pl,aie} or with a pion
exchange potential \cite{olof} (however in the latter case no
$E2$ transition calculation is available). At the photon point, the
results on the
quadrupole $N-\Delta$ transition are given in terms of the ratio

\begin{equation} 
R�=�- \frac{G_{E2}}{G_{M1}} 
\end{equation} 
where $G_{E2}$ and
$G_{M1}$ are the transverse electric and magnetic transition strengths,
respectively. The PDG value is \cite{pdg} $R�=�-0.02 \pm 0.005$. A number
not far from this was obtained with the CQM including a hyperfine
interaction
\cite{ki,ikk}. In particular, taking care of the higher shells and of the
Siegert's theorem for a more accurate and reliable calculation, the value
$R=� � 0.02$ was obtained \cite{dg}. However one should not forget that
the $M1$ transition is underestimated and then, even if the ratio is
correctly reproduced, the quadrupole strength still remains too low.
An estimate of the total quadrupole excitation strength for the nucleon
can
be obtained from an energy weighted sum rule approach to the excitation of
the quark
degrees of freedom \cite{briz}; the $F_{15}(1680)$ resonance,
which is mainly a D-state, saturates only $30\%$ of the sum rule,
showing that the missing strength is expected to be spread over the whole
spectrum. 

The inadequacy of CQMs to reproduce the quadrupole photon excitation is
visible also when one studies the $Q^2$ behaviour of the transition form
factors. In this case also the longitudinal amplitude $S_{1/2}$ must be
considered. The situation is illustrated in Fig. 1 \cite{ts03}, where the
longitudinal
form factor for the $N-\Delta$ transition, calculated with the hCQM, is
reported in comparison with a global fit performed by the Mainz group
\cite{maid}. The theoretical $S_{1/2}$ is very small in comparison with
data.

\begin{figure}[h]
\includegraphics[width=7cm]{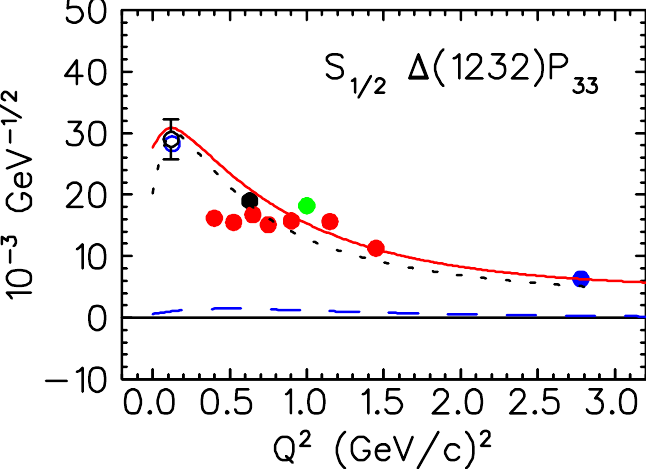}
%\centerline{\epsfig{file=a12.ps,width=6.0cm,angle=0}}
%\vspace{0.3cm} \centerline{\epsfig{file=a32.ps,width=6.0cm,angle=0}}
%\vspace{0.3cm} \centerline{\epsfig{file=s12.ps,width=6.0cm,angle=0}}
\caption{ The $Q^2$ dependence of the $N \rightarrow \Delta$
longitudinal helicity
amplitude $S_{1/2}$: superglobal fit performed with MAID \cite{maid}
(solid curve),
predictions of the hypercentral Constituent Quark Model
\cite{pl,ts03,helamp} (dashed curve), pion cloud contributions calculated
with the Mainz dynamical model \cite{DMT} (dotted curve). The data points
at finite $Q^2$ are the results of single-Q$^2$ fits \cite{ts03} on recent
data \cite{dat1,dat2,dat3,dat4,dat5,dat6}. }
\end{figure}

A similar underestimate occurs also for the transverse helicity
amplitudes $A_{1/2},A_{3/2}$ (see Figs. 2 and 3) and therefore also for
the 
the transverse electric $G_{E2}$ and the magnetic $G_{M1}$ from factors,
since by definition they are proportional to linear combinations of the
transverse helicity amplitudes:
\begin{equation}
G_{E2} �\propto���A_{3/2}�- \sqrt{3}�A_{1/2}
\end{equation}
\begin{equation}
G_{M1} �\propto���\sqrt{3}�A_{3/2}�+ A_{1/2}
\end{equation}

\begin{figure}[htb]
\includegraphics[height=.3\textheight]{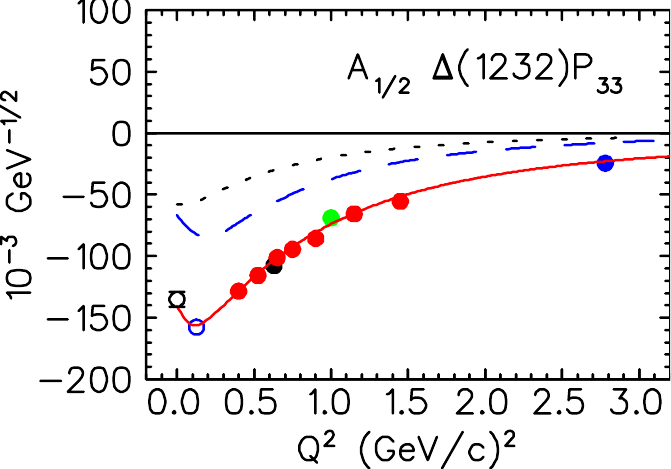}
\caption{ The same as in Fig. 1 but for the transverse helicity
amplitude $A_{1/2}$. At $Q^2=0$ the photon
coupling from PDG is shown \cite{pdg}.}
\end{figure}

\begin{figure}[htb]
\includegraphics[height=.3\textheight]{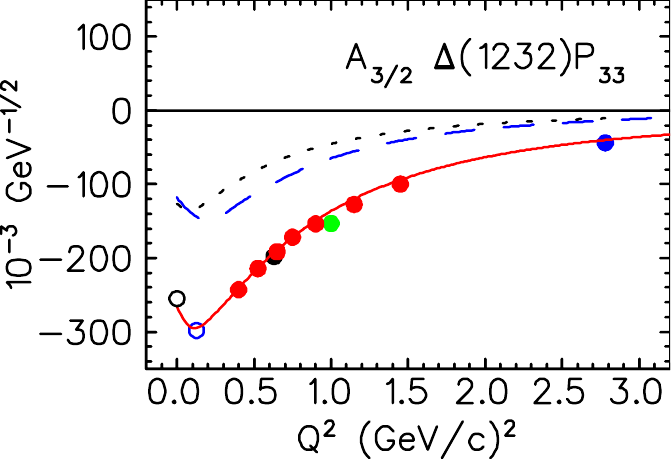}
\caption{ The same as in Fig. 1 but for the transverse helicity
amplitudes $A_{3/2}$. At $Q^2=0$ the photon
coupling from PDG is shown \cite{pdg}.}
\end{figure}

The
seriousness
of this discrepancy is enhanced by the expectation that, on the basis of
helicity conservation in the virtual photon-quark interaction, the ratio
\begin{equation} 
A�=�\frac{|A_{1/2}|^2�-�|A_{3/2}|^2}{|A_{1/2}|^2�+�|A_{3/2}|^2} 
\end{equation} 
is expected to reach the
value $1$ \cite{carl} if $Q^2$ goes to infinity. It is easy to show that
\begin{equation} 
A�=�-\frac{1}{2}�+�\frac{3�G_{E2}�(G_{E2}�-�G_{M1})}{G_{M1}^2�+�3� G_{E2}^2} 
\end{equation} 
which implies that the value $A=1$
is reached for $ G_{E2}�=�- G_{M1}$.

A possible reason of this discrepancy can be envisaged looking again at
Figs. 1, 2 and 3, where the contributions of the meson cloud \cite{DMT} to
the $\Delta$ helicity amplitudes are reported. But this point will be
discussed in the next section.

\section{Meson and Quark pair effects}

As quoted in the last section, the various models reproduce the overall
trend of the photocouplings, in particular the cases where the
excitation strength is vanishing. It should be reminded that in the
earlier h.o. calculations \cite{cko} the vanishing of the $A_{1/2}$
helicity
amplitude for the excitation of the proton to the  $F_{15}(1680)$
resonace was obtained imposing the
proton radius to be of the order of $0.5�fm$. In this way also the
$A_{1/2}$ helicity
amplitude for the proton transition to the $D_{13}(1520)$ turns out to be
small. It is worthwhile noting that the calculated proton
radius in the hCQM is actually about $0.5�fm$; this fact is one of the
reasons why the
hCQM predictions for the helicity amplitudes of the negative parity
resonances are in reasonable agreement with data \cite{aie2}.

The smallness of the proton radius required for the description of the
e.m. excitation together with the lack of strength in the low $Q^2$
region, suggest an interesting picture for the proton
\cite{es,aie,aie2} (and consequently
for hadrons), namely that of a small core, with radius of about
$0.5�fm$, surrounded by an external cloud made of mesons
and/or quark-antiquark pairs. The
contributions coming from this external cloud have been pointed out as
a possible origin of the missing strength 
\cite{es,aie,aie2} and
are obviously lacking in the 
available
CQMs; their effect is expected to decrease for medium-high $Q^2$ and
therefore it is not a surprise that the hCQM fails to reproduce the
strength at
the photon point but give reasonable results for medium $Q^2$.  These
considerations are supported by the inspection of Fig. 1, 2 and 3, where
the pion cloud contributions, evaluated by means of a dynamical model 
 \cite{DMT}, are
reported. Their importance decreases with increasing $Q^2$,
going rapidly to zero, as expected. This feature is quite general, since
it happens systematically also for the excitation of higher resonances,
such as $P_{11}(1440), S_{11}(1535), D_{13}(1520), F_{15}(1680)$
\cite{ts03}. It is important to note that the pion contributions tend to
fill the gap between the pure valence quark calculations and the data. In
particular, the quadrupole strength observed in the $N-\Delta$ excitation
seems to be substantially due to meson effects or, stated in another way,
to sea quark effects. Therefore the shape of baryons is determined not
only by the quark core but also by the meson or quark-antiquark pair
cloud.

The quark-antiquark pair and/or meson cloud effects are relevant in many
properties of hadrons. One important case is the width of resonances,
coming from the decay of baryons in the meson nucleon channel. This implies a
coupling with the continuum,
which is not taken into account in the present CQMs: a quark-meson
vertex is in some models introduced in order to describe the strong
decay \cite{ki,cr3,cr,bil3,melde}, but the
theoretical baryon states are all with zero width. The spectrum
itself presents some features which might be a manifestation of quark
pair effects, namely the isospin dependence which is necessary to describe
the position of some states such as the Roper. This is clear in the 
GBE model \cite{olof}, where a pion is explicitly
exchanged, but also in other models \cite{bil,iso} the isospin
dependence may be a remnant of quark-antiquark effects. However, 
an explicit
manifestation of quark pair and/or meson effects should be looked for in
the baryon widths. A work on these lines has been done some time
ago: the IK model has been extended introducing a direct quark-meson
coupling by means of appropriate interaction lagrangians and used for
the description of both masses and widths of baryons
\cite{blask}. 

The elastic nucleon form factors provide another example of physical
quantities for which such effects are expected to be
relevant. Because of the smallness of
the quark core radius, the non relativistic calculations are not able to
reproduce the nucleon form factor data. The inclusion of first order  
relativistic corrections generated by Lorentz boosts gives rise to a
substantial improvement \cite{mds} and, moreover, it has been shown to
lead to a decrease of the ratio $G_E/G_M$ between the electric and
magnetic proton form
factors \cite{rap}, in qualitative agreement with the recent Jlab
measurements \cite{Jlab}. 

For a good description of the nucleon elastic
form factors a completely relativistic approach is needed. The
internal quark motion is usually described in the baryon rest frame, but
the elastic form factors are evaluated in the Breit frame, 
in motion with respect to both the initial and final nucleon rest frame
with a velocity which increases with the virtual photon momentum transfer 
$Q^2$. Therefore, the theory should be formulated in a covariant way, 
transforming correctly the three quark states by means of Lorentz
boosts and making use of a covariant quark current. Various relativistic
calculations of the elastic nucleon form factors are now available
\cite{card_ff,wagen,boffi,mert,ff_05},obtaining a good description of
data.
However, in order to achieve a detailed account of the experimental
behaviour \cite{ff_05}, in
particular
of the decrease of the $G_E/G_M$ ratio and of the small $Q^2$ 
wiggles in the proton form factors, one has to
introduce intrinsic
quark form factors. Actually constituent quarks are effective degrees of
freedom, which take implicitly into account complicated quark-gluon
 interactions, which of course contain also quark pair effects.
(For a review concerning the elastic nucleon form factors the reader is
referred to \cite{dejager}).

The inclusion of meson effects in hadron properties is now receiving
considerable attention and in this Workshop there are numerous examples.
Concerning the electromagnetic excitation of nucleon resonances we can
quote the Mainz dynamical model \cite{DMT}, the coupled channel approach
\cite{lee}
 and the inclusion of explicit 3q-pion components in the nucleon
state \cite{dong}.

\section{Conclusion}

The study of the electromagnetic excitation of the nucleon resonances
offers the opportunity for a sensitive test of the CQMs. The Nucleon-$\Delta$ 
case is particularly interesting because the $\Delta$ resonance is  
easily and strongly excited and its internal structure is
very similar to
that of the nucleon. Moreover, some models, namely those which
consider a tensor-like force between quarks, predict a deformation of both
the nucleon and the $\Delta$, a deformation which can manifests itself in
a longitudinal and transverse quadrupole excitation strength. CQM
calculations, performed consistently with other baryon properties, in
particular the spectrum, predict strengths which are very low in
comparison with the observed ones. On the other hand, the pion
contributions, evaluated by means of a dynamical model, have been shown to
be able to fill. at least partially, the gap between data and theoretical
predictions, supporting a view of the nucleon as a small quark
core surrounded by an external meson (or quark pair) cloud. This means in
particular that the shape of hadrons is to a large extent determined by
such meson cloud effects, which will be certainly object of intense
studies in the near future.

An important issue connected with the study of the internal baryon
structure is provided by relativiy. This means first of all the necessity
of introducing the relativistic kinetic
energy in order to describe correctly the internal quark motion, 
also in case of small constituent quark masses. More relevant is the
formulation of the model within a consistent relativistic framework, which
means a relativistic hamiltonian in any of the allowed forms
(light front, instant or point) or a Bethe-Salpeter approach. The
inclusion of relativity, specially in the sense of considering a covariant
quark current, is crucial for the description of the elastic form factors.
However, for the electromagnetic transition
form factors,
the relativistic corrections, at least at the first order level, are
meaningful but not determinant \cite{mds2}. On the other hand, the
spectrum seems to be not
sensitive to relativity, provided that the quark masses are not too low.

The analysis of the theoretical predictions concerning the $Q^2$ behaviour
of the helicity amplitudes in general, but in particular in the case of
the $N-\Delta$ transition, shows that some fundamental mechanism is
lacking and there are indications that meson cloud and/or quark-antiquark
pair effects should be included in the CQM desscription of hadron
properties.

\end{document}